\documentclass[journal=jcp,manuscript=article]{achemso}
\setkeys{acs}{maxauthors=0}

\usepackage{achemso}
\usepackage{graphics}
\usepackage{amssymb,amsfonts}
\usepackage{graphicx}
\usepackage[table]{xcolor}
\usepackage{caption}
\usepackage{subcaption}
\usepackage{colortbl}
\usepackage{amsmath}
\usepackage{amsopn}
\usepackage{bm}
\usepackage{color}
\usepackage{array}
\usepackage{lscape}
\usepackage{mciteplus}
\usepackage[version=3]{mhchem}
\usepackage{ulem}
\usepackage{listings}
\usepackage{enumerate}
\SectionNumbersOn
\makeatletter
\newcommand*\rel@kern[1]{\kern#1\dimexpr\macc@kerna}
\newcommand*\widebar[1]{%
  \begingroup
  \def\mathaccent##1##2{%
    \rel@kern{0.8}%
    \overline{\rel@kern{-0.8}\macc@nucleus\rel@kern{0.2}}%
    \rel@kern{-0.2}%
  }%
  \macc@depth\@ne
  \let\math@bgroup\@empty \let\math@egroup\macc@set@skewchar
  \mathsurround\z@ \frozen@everymath{\mathgroup\macc@group\relax}%
  \macc@set@skewchar\relax
  \let\mathaccentV\macc@nested@a
  \macc@nested@a\relax111{#1}%
  \endgroup
}
\makeatother
\numberwithin{equation}{subsection}

\makeatletter
\renewcommand*\env@matrix[1][\arraystretch]{%
  \edef\arraystretch{#1}%
  \hskip -\arraycolsep
  \let\@ifnextchar\new@ifnextchar
  \array{*\c@MaxMatrixCols c}}
\makeatother
\usepackage{rotating}
\usepackage{pifont}% http://ctan.org/pkg/pifont
\def\ket#1{|#1\rangle}
\def\bra#1{\langle#1|}

\newcommand{\hf}{ \mathrm{HF} }

\newcommand{\Ts}{ {^{*}}\hat{T}}
\newcommand{\ts}{ {^{*}}t}

\newcommand{\ECCthree}{\text{E-CCSD(T--3)}}
\newcommand{\ECCfour}{\text{E-CCSD(T--4)}}

\newcommand{\ECCseven}{\text{E-CCSD(T--7)}}
\newcommand{\ECCn}{\text{E-CCSD(T--$n$)}}

\newcommand{\CCtwo}{\text{CCSD(T--2)}}
\newcommand{\CCthree}{\text{CCSD(T--3)}}
\newcommand{\CCfour}{\text{CCSD(T--4)}}

\newcommand{\CCn}{\text{CCSD(T--$n$)}}

\newcommand{\CCSD}{\text{CCSD}}
\newcommand{\CCSDT}{\text{CCSDT}}
\newcommand{\HSDT}{e^{-\delta\hat{T}} \hat{H}^{\Ts} e^{\delta\hat{T}}}
\newcommand{\HSDTz}{e^{-\delta\hat{T}} \hat{H}^{T^{(0)}} e^{\delta\hat{T}}}
\newcommand{\HSD}{ \hat{H}^{\Ts} }
\newcommand{\PhiSD}{ \hat{\Phi}^{\Ts} }
\newcommand{\tb}{\bar{t}}
\newcommand{\tbe}{\delta \bar{t}^E}
\newcommand{\tbeone}{\delta \bar{t}^{E(1)}}
\newcommand{\tbetwo}{\delta \bar{t}^{E(2)}}

\newcommand{\tbl}{\delta \bar{t}^L}
\newcommand{\tblone}{\delta \bar{t}^{L(1)}}

\newcommand{\tbs}{ {^{*}}\bar{t}}
\newcommand{\tsbold}{{^{*}}\mathbf{t}}
\newcommand{\tzbold}{\mathbf{t}^{(0)}}
\newcommand{\tbsbold}{{^{*}}\mathbf{\bar{t}}}
\newcommand{\tbzbold}{ \mathbf{\bar{t}}^{(0)} }
\newcommand{\dtbold}{\delta\mathbf{t}}
\newcommand{\dtbbold}{\delta\mathbf{\bar{t}}}
\newcommand{\dtbbolde}{\delta\mathbf{\bar{t}}^E}
\newcommand{\dtbboldl}{\delta\mathbf{\bar{t}}^L}
\newcommand{\zbold}{\mathbf{0}}

\newcommand{\tbz}{\bar{t}^{(0)}}

\author{Kasper Kristensen}
\email{kasperk@chem.au.dk}
\affiliation[Aarhus University]
{qLEAP Center for Theoretical Chemistry,
Department of Chemistry,
Aarhus University, Langelandsgade 140, DK-8000 Aarhus C, Denmark}
\author{Janus Juul Eriksen}
\email{janusje@chem.au.dk}
\affiliation[Aarhus University]
{qLEAP Center for Theoretical Chemistry,
Department of Chemistry,
Aarhus University, Langelandsgade 140, DK-8000 Aarhus C, Denmark}
\author{Devin A. Matthews}
\affiliation[The University of Texas at Austin]
{The Institute for Computational Engineering and Sciences, The University of Texas at Austin, Austin, Texas 78712, USA}
\author{Jeppe Olsen}
\affiliation[Aarhus University]
{qLEAP Center for Theoretical Chemistry,
Department of Chemistry,
Aarhus University, Langelandsgade 140, DK-8000 Aarhus C, Denmark}
\author{Poul J{\o}rgensen}
\affiliation[Aarhus University]
{qLEAP Center for Theoretical Chemistry,
Department of Chemistry,
Aarhus University, Langelandsgade 140, DK-8000 Aarhus C, Denmark}

\title[TITLE]
  {A view on coupled cluster perturbation theory using a bivariational Lagrangian formulation}

\begin{document}
%

%
%%%%%%%%%%%%%%%%%%%%%%%%%%%%%%%%%%%%%%%%%%%%%%%%%%%%%%%%%%%%%%%%%%%
%                                                                    				Abstract
%%%%%%%%%%%%%%%%%%%%%%%%%%%%%%%%%%%%%%%%%%%%%%%%%%%%%%%%%%%%%%%%%%%
%
\begin{abstract}
We consider two distinct coupled cluster (CC) perturbation series that both expand the difference between the energies of the CCSD (CC with single and double excitations) and CCSDT (CC with single, double, and triple excitations) models in orders of the M{\o}ller-Plesset fluctuation potential. We initially introduce the $\ECCn$ series, in which the CCSD amplitude equations are satisfied at the expansion point, and compare it to the recently developed $\CCn$ series [J. Chem. Phys. $\bm{140}$, 064108 (2014)], in which not only the CCSD amplitude, but also the CCSD multiplier equations are satisfied at the expansion point. The computational scaling is similar for the two series, and both are term-wise size extensive with a formal convergence towards the CCSDT target energy. However, the two series are different, and the $\CCn$ series is found to exhibit a more rapid convergence up through the series, which we trace back to the fact that more information at the expansion point is utilized than for the $\ECCn$ series. The present analysis can be generalized to any perturbation expansion representing the difference between a parent CC model and a higher-level target CC model. In general, we demonstrate that, whenever the parent parameters depend upon the perturbation operator, a perturbation expansion of the CC energy (where only parent amplitudes are used) differs from a perturbation expansion of the CC Lagrangian (where both parent amplitudes {\it and} parent multipliers are used). For the latter case, the bivariational Lagrangian formulation becomes more than a convenient mathematical tool, since it facilitates a different and faster convergent perturbation series than the simpler energy-based expansion.

\end{abstract}
\newpage
%

%
%%%%%%%%%%%%%%%%%%%%%%%%%%%%%%%%%%%%%%%%%%%%%%%%%%%%%%%%%%%%%%%%%%%
%		                                                 					 Introduction
%%%%%%%%%%%%%%%%%%%%%%%%%%%%%%%%%%%%%%%%%%%%%%%%%%%%%%%%%%%%%%%%%%%
%
\section{Introduction}\label{sec:intro}

Coupled cluster (CC) theory~\cite{cicek_1,cicek_2,paldus_cikek_shavitt,shavitt_bartlett_cc_book} is perhaps the most powerful method for describing dynamical electron correlation effects within the realm of modern quantum chemistry. The CC singles and doubles (CCSD) model~\cite{ccsd_paper_1_jcp_1982}, in which the cluster operator is truncated at the level of double excitations, is a robust and useful model, but it is well-known that the effects of triple (and higher-level) excitations need to be taken into account in order to obtain highly accurate results that may compete with the accuracy of experiments~\cite{mest}. However, the steep scaling of the CC singles, doubles, and triples~\cite{ccsdt_noga_bartlett_jcp_1987,ccsdt_paper_2_cpl_1988} (CCSDT) and CC singles, doubles, triples, and quadruples~\cite{ccsdtq_paper_1_jcp_1991,ccsdtq_paper_2_jcp_1992} (CCSDTQ) models limits their use to rather modest molecular systems. For this reason, a computationally tractable alternative to the expensive iterative CCSDT and CCSDTQ models is to develop approximate models, for which the important triples and/or quadruples contributions are determined from a perturbation analysis, and hence included in a cheap and non-iterative fashion. A plethora of different models for the approximate treatment of triples and/or quadruples effects have been suggested, and we refer to Ref. \citenum{eriksen2014lagrangian} for a recent theoretical overview of approximate non-iterative triples and quadruples models and Refs. \citenum{eriksen2015convergence} and \citenum{quadruples_pert_theory_jcp_2015} for a numerical comparison of many of these.

In the present work, we focus on perturbation theory within a CC framework, where a M{\o}ller-Plesset (MP) partitioning of the Hamiltonian is performed~\cite{MP}, and the energy difference between a zeroth-order (parent) CC model and a higher-level (target) CC model is expanded in orders of the perturbation (the fluctuation potential). In particular, we will base the perturbation analysis on a bivariational CC Lagrangian obtained by adding to the CC target energy the CC amplitude equations with associated Lagrange multipliers. We note that the linearly parametrized state formally spanned by the Lagrange multipliers is often referred to as the CC $\Lambda$-state~\cite{handy_schaefer_lambda_ci_jcp_1984,schaefer_lambda_cc_jcp_1987}, and that this is in general different from the exponentially parametrized CC state. As pointed out by Arponen~\cite{arponen_ann_phys_1983,arponen_pra_1987}, extensively exploited in the CC Lagrangian formulation\cite{helgaker_jorgensen_lagrangian_aqc_1988}, and recently discussed by Kvaal~\cite{kvaal_jcp_2012,kvaal_mol_phys_2013}, the CC energy may be interpreted as a CC functional in both the CC amplitudes and the $\Lambda$-state parameters. We will show that the fastest convergence is obtained when these two sets of state parameters are treated on an equal footing in the perturbative expansion of the energy difference between a parent and a target CC model. Thus, we will distinguish between a perturbation expansion of the CC energy, for which only parent amplitudes are used at the expansion point, and a perturbation expansion of the CC Lagrangian, for which both parent amplitudes {\it and} parent multipliers are used at the expansion point. At first sight, the bivariational Lagrangian formulation might seem like an unnecessary complication, since, for the target model, the Lagrangian formally equals the energy. The purpose of this work is, however, to highlight and exemplify that not only is the Lagrangian formulation a convenient mathematical tool that may simplify the derivation of various perturbation expansions; in many cases, the Lagrangian formulation will actually lead to different perturbation series than the corresponding energy formulation. 

To exemplify this difference, we consider two perturbation series which both expand the difference between the energies of the CCSD and CCSDT models. We initially introduce the $\ECCn$ series, in which the CCSD amplitude equations are satisfied at the expansion point, and next compare it to the recently developed $\CCn$ series~\cite{eriksen2014lagrangian}, in which not only the CCSD amplitude equations are satisfied at the expansion point, but also the CCSD multiplier equations. Despite depending on the fluctuation potential to infinite order in the space of all single and double excitations, the CCSD amplitudes are formally considered as zeroth-order parameters in both the $\ECCn$ and $\CCn$ series, since the CCSD model represents the expansion point. Similarly, the CCSD multipliers, which too depend on the fluctuation potential, are considered as zeroth-order parameters for the $\CCn$ series, but not so for the $\ECCn$ series. The path from the CCSD expansion point towards the CCSDT target energy, as defined by a perturbation expansion, is thus different within the $\ECCn$ and $\CCn$ series, and, as will be shown in the present work, the $\CCn$ series is the more rapidly converging of the two, since more information is utilized at the expansion point. We finally reiterate that the lowest-order contribution of the $\CCn$ series (that of the $\CCtwo$ model) is identical to the triples-only part of the CCSD(2) model of Gwaltney and Head-Gordon~\cite{ccsd_2_model_gwaltney_head_gordon_cpl_2000,ccsd_2_model_gwaltney_head_gordon_jcp_2001} and the second-order model of the CC($2$)PT($m$) series of Hirata {\it et al.}~\cite{ccsd_pt_models_hirata_jcp_2001,ccsd_pt_models_hirata_jcp_2004,ccsd_pt_models_hirata_jcp_2007}, the $\CCthree$ model is identical to the triples-only part of the third-order CC($2$)PT($m$) model, while for fourth and higher orders, the $\CCn$ and CC($2$)PT($n$) series are different~\cite{eom_cc_pert_theory_jcp_2014}. The $\ECCn$ series, however, differs from the aforementioned perturbation series for all corrections.\\

The present study is outlined as follows. In \ref{sec:theory}, we derive the $\ECCn$ series and compare it to the $\CCn$ series in order to illustrate the importance of treating parent amplitudes and multipliers on an equal footing. In \ref{sec:results}, we present numerical results for the $\ECCn$ and $\CCn$ energies, while some concluding remarks are given in \ref{sec:conclusion}.

%
%%%%%%%%%%%%%%%%%%%%%%%%%%%%%%%%%%%%%%%%%%%%%%%%%%%%%%%%%%%%%%%%%%%
%		                                                 					 Theory
%%%%%%%%%%%%%%%%%%%%%%%%%%%%%%%%%%%%%%%%%%%%%%%%%%%%%%%%%%%%%%%%%%%
%
\section{Theory}\label{sec:theory}

In this section, we consider two perturbation series that expand the difference between the CCSD and CCSDT energies in orders of the perturbation. In \ref{sec:eccsdt}, we develop a new energy-based perturbation series denoted the $\ECCn$ series, which can be formulated in terms of CC amplitudes without the need for invoking a CC Lagrangian. Next, in \ref{sec:general}, we develop a common bivariational framework, in which we recast the $\ECCn$ and $\CCn$ series. Finally, we present a theoretical comparison between the two series in \ref{sec:comparison} in order to exemplify how the CC Lagrangian framework may lead to a perturbation series that is inherently different from that which arise from the corresponding energy formulation.

%
%%%%%%%%%%%%%%%%%%%%%%%%%%%%%%%%%%%%%%%%%%%%%%%%%%%%%%%%%%%%%%%%%%%
%		                                                  Perturbation expansion based on the CCSDT energy
%%%%%%%%%%%%%%%%%%%%%%%%%%%%%%%%%%%%%%%%%%%%%%%%%%%%%%%%%%%%%%%%%%%
%
\subsection{Perturbation expansion based on the CCSDT energy}\label{sec:eccsdt}

In this work, we use a MP partitioning of the Hamiltonian
\begin{align}
\label{HFPhi}
\hat{H} = \hat{f} + \hat{\Phi}
\end{align}
where $\hat{f}$ is the Fock operator and $\hat{\Phi}$ is the fluctuation potential. We consider first the CCSD model, which we choose as the common reference point for the perturbation expansions to follow. The CCSD energy, $E^{\CCSD}$, and associated amplitude equations may be written as
\begin{align}
\label{ETs}
E^{\CCSD}  = \bra{\hf} \hat{H}^{\Ts} \ket{\hf}
\end{align}
and
\begin{align}
\label{CCSDamp}
0 = \bra{\mu_i} \hat{H}^{\Ts} \ket{\hf}
\qquad (i=1,2)
\end{align}
where $\bra{\mu_1}$ and $\bra{\mu_2}$ represent a singly and a doubly excited state with respect to the HF determinant, $\ket{\hf}$, and the (non-Hermitian) CCSD similarity-transformed Hamiltonian is given by
\begin{align}
\label{HTs}
\hat{H}^{\Ts} &= e^{- \Ts } \hat{H} e^{ \Ts}, 
\qquad
\Ts = \Ts_1 + \Ts_2
\end{align}
with $\Ts_1$ and $\Ts_2$ being the CCSD singles and doubles cluster operators. Throughout the paper, we will use asterisks to denote CCSD quantities and generally use the generic notation $\hat{T}_i = \sum_{\mu_i} t_{\mu_i} \hat{\tau}_{\mu_i}$ for a cluster operator at excitation level $i$, where $\hat{\tau}_{\mu_i}$ is an excitation operator and $t_{\mu_i}$ is the associated amplitude.

We now parametrize the difference between the CCSD and CCSDT energy in terms of correction amplitudes, $\delta t_{\mu_i}$, which represent the difference between the CCSD and CCSDT amplitudes. The correction amplitudes are expanded in orders of the fluctuation potential, and the CCSDT amplitudes, $t_{\mu_i}$, may thus be written as
\begin{align}
\label{CCSDT_CCSD_exp}
t_{\mu_i} = t_{\mu_i}^{(0)} + \delta t_{\mu_i}^{(1)} + \delta t_{\mu_i}^{(2)} + \ldots \qquad (i=1,2,3)
\end{align}
where $t_{\mu_i}^{(0)} = \ts_{\mu_i}$ for $i=1,2$ and $t_{\mu_3}^{(0)} = 0$. We emphasize that, since we have chosen to expand the CCSDT amplitudes around the CCSD reference point, the CCSD amplitudes, $\{\ts_{\mu_1},\ts_{\mu_2}\}$, are zeroth-order by definition. The $\{\delta t_{\mu_1},\delta t_{\mu_2}\}$ amplitudes thus represent corrections to the CCSD amplitudes, while $\{\delta t_{\mu_3}\}$ are the CCSDT triples amplitudes.

The CCSDT cluster operator may now be written as $\hat{T} = \Ts + \delta\hat{T}$, where $\delta\hat{T}$ contains the correction amplitudes, and the CCSDT energy may be obtained by projecting the CCSDT Schr{\"o}dinger equation, $\HSDT \ket{\hf} = E^{\CCSDT} \ket{\hf}$, against $\bra{\hf}$
\begin{align}
\nonumber
E^{\CCSDT}  &= \bra{\hf} \HSDT \ket{\hf} \\
&= 
E^{\CCSD} + \bra{\hf} [ \PhiSD, \delta\hat{T}_1 + \delta\hat{T}_2] + \tfrac{1}{2} [[\PhiSD, \delta\hat{T}_1],\delta\hat{T}_1] \ket{\hf}
\label{ETf3}
\end{align}
where we have carried out a Baker-Campbell-Hausdorff (BCH) expansion, while the CCSDT amplitude equations are obtained by projection against the combined excitation manifold of all single, double, and triple excitations out of the HF reference
\begin{align}
\label{ampeq}
 0 = \bra{\mu_i} \HSDT \ket{\hf} 
\qquad (i=1,2,3) \ .
\end{align} 
The order analysis of \ref{ampeq} is identical to the one performed in Ref. \citenum{eriksen2014lagrangian} (orders are counted in $\hat{\Phi}$), and the resulting amplitudes are thus the same. Compactly, these are given by
\begin{subequations}
\label{amp_pert_expand}
\begin{align}
\label{ampeqSD}
\delta t_{\mu_i}^{(n)} &= 
- \epsilon_{\mu_i}^{-1} \bigg(
 \bra{\mu_i} [\PhiSD, \delta\hat{T}] 
+ \tfrac{1}{2} [ [\PhiSD, \delta\hat{T}] , \delta\hat{T}] + \ldots \ket{\hf} 
\bigg)^{(n)}
\qquad (i=1,2)
\\
\delta t_{\mu_3}^{(n)} &= 
- \epsilon_{\mu_3}^{-1} \bigg(
 \bra{\mu_3} \PhiSD + [\PhiSD, \delta\hat{T}] 
+ \tfrac{1}{2} [ [\PhiSD, \delta\hat{T}] , \delta\hat{T}] + \ldots \ket{\hf} 
\bigg)^{(n)}
\end{align}
\end{subequations}
where $\epsilon_{\mu_i}$ is the orbital energy difference between the virtual and occupied spin-orbitals of excitation $\mu_i$, and the right-hand sides of the equations contain all terms of order $n$, i.e., the sum of the orders of all $\delta T$ operators plus one (for the fluctuation potential) equals $n$. For example, the first-order singles and doubles corrections are zero, $\delta t_{\mu_1}^{(1)} = \delta t_{\mu_2}^{(1)} = 0$, while the first-order triples corrections are given as  $\delta t_{\mu_3}^{(1)} =- \epsilon_{\mu_3}^{-1}  \bra{\mu_3} \PhiSD  \ket{\hf}$. By collecting terms in orders of the fluctuation potential, the CCSDT energy in \ref{ETf3} may now be expanded as
\begin{subequations}
\label{ECCSDTn}
\begin{align}
E^{\CCSDT} &= 
E^{\CCSD} + \sum_{n=3}^{\infty} E^{(n)}
\\
\nonumber
E^{(n)} &=  \bra{\hf} [ \PhiSD, \delta\hat{T}_1^{(n-1)} + \delta\hat{T}_2^{(n-1)}] \ket{\hf} 
\\
&+ 
\tfrac{1}{2} \sum_{m=2}^{n-3} \bra{\hf} [[\PhiSD, \delta\hat{T}_1^{(m)}], \delta\hat{T}_1^{(n-m-1)}] \ket{\hf}
\label{ECCSDTn2}
\end{align}
\end{subequations}
where we have used the fact that the first-order singles and doubles amplitudes vanish to restrict the summations. We denote the perturbation series defined by \ref{ECCSDTn} as the {\it $\ECCn$ series} to emphasize that it is based on a perturbation expansion of the CCSDT energy around the CCSD {\it energy} point, at which the CCSD amplitude equations are satisfied. We note that this notation is not to be confused with ECC, which is usually used as an acronym for extended coupled cluster theory in the literature~\cite{arponen_ann_phys_1983,arponen_pra_1987}

From \ref{ECCSDTn}, it follows that the first non-vanishing energy correction to the $\ECCn$ series is of third order. The two lowest-order corrections are given in \ref{E3alt_app_n_1} and \ref{E4alt_app_n_1} of \ref{appendix_comparison}. The $\ECCn$ series is evidently different from the $\CCn$ series developed in Ref. \citenum{eriksen2014lagrangian}, which starts at second order. Both series, however, describe the difference between the CCSD and CCSDT energies using a MP partitioning of the Hamiltonian, and the correction amplitudes are identical. The only apparent difference is that the CCSDT energy is the central quantity for the $\ECCn$ series, while the CCSDT Lagrangian forms the basis for the $\CCn$ series. In \ref{sec:general}, we develop a general Lagrangian framework to enable a direct comparison of the $\ECCn$ and $\CCn$ series in \ref{sec:comparison}.

%
%%%%%%%%%%%%%%%%%%%%%%%%%%%%%%%%%%%%%%%%%%%%%%%%%%%%%%%%%%%%%%%%%%%
%		                                                  Comparison of energy- and Lagrangian-based perturbation theory
%%%%%%%%%%%%%%%%%%%%%%%%%%%%%%%%%%%%%%%%%%%%%%%%%%%%%%%%%%%%%%%%%%%
%
\subsection{A general bivariational Lagrangian framework}\label{sec:general}

The CCSDT Lagrangian may be obtained by adding to the CCSDT energy the amplitude equations in \ref{ampeq} with associated (undetermined) multipliers
\begin{align}
\label{LT1}
L^{\CCSDT}(\tzbold,\tbzbold,\dtbold,\dtbbold) = E^{\CCSDT}
+ \sum_{j=1}^3 \sum_{\nu_j} ( \tbz_{\nu_j} + \delta \tb_{\nu_j} ) \bra{\nu_{j}} \HSDTz  \ket{\hf}
\end{align}
where we have chosen the following parametrization of the CCSDT multipliers in analogy with \ref{CCSDT_CCSD_exp}
\begin{align}
\label{multpara}
\tb_{\mu_i} = \tb_{\mu_i}^{(0)} + \delta \tb_{\mu_i}^{(1)} + \delta \tb_{\mu_i}^{(2)} + \ldots \qquad (i=1,2,3) \ .
\end{align}
If the expansion of the CCSDT multipliers in \ref{multpara} is left untruncated, these will equal the parameters of the linearly parametrized CCSDT $\Lambda$-state. The notation for the Lagrangian, $L^{\CCSDT}(\tzbold,\tbzbold,\dtbold,\dtbbold)$, highlights that the amplitudes and multipliers at the expansion point are $\tzbold$ and $\tbzbold$, respectively, while $\dtbold$ and $\dtbbold$ represent the correction amplitudes and correction multipliers, respectively. By setting $\tzbold = \tbzbold = \zbold$, we arrive at an MP-like perturbation expansion, albeit one that has the CCSDT energy as the target energy instead of the full configuration interaction (FCI) energy. In this work, however, we focus on the case for which the CCSD amplitudes are used as zeroth-order amplitudes ($t_{\mu_1}^{(0)} = \ts_{\mu_1}$, $t_{\mu_2}^{(0)} = \ts_{\mu_2}$, $t_{\mu_3}^{(0)} = 0$), while considering different choices of zeroth-order multipliers. In particular, we show below how the $\ECCn$ series may be recovered by choosing $\tb_{\mu_i}^{(0)} = 0$ (i=1,2,3), while the $\CCn$ series corresponds to using CCSD multipliers as zeroth-order multipliers
($\tb_{\mu_1}^{(0)} = \tbs_{\mu_1}$, $\tb_{\mu_2}^{(0)} = \tbs_{\mu_2}$, $\tb_{\mu_3}^{(0)} = 0$). Equations for the CCSD multipliers are obtained by requiring the CCSD Lagrangian
\begin{align}
L^{\CCSD}  = \bra{\hf} \HSD \ket{\hf} + \sum_{j=1}^2 \sum_{\nu_j} \tbs_{\nu_j} \bra{\nu_{j}} \HSD  \ket{\hf}
\label{LCCSD}
\end{align}
to be stationary with respect to variations in the CCSD amplitudes
\begin{align}
\label{CCSDmult}
0 = \frac{\partial L^{\CCSD}}{\partial \ts_{\mu_i}} = 
\bra{\hf} [ \PhiSD, \hat{\tau}_{\mu_i} ] \ket{\hf}
+ \sum_{j=1}^2 \sum_{\nu_j} \tbs_{\nu_j}
\bra{\nu_j} [ \HSD, \hat{\tau}_{\mu_i} ] \ket{\hf}
\qquad
(i=1,2) \ .
\end{align}

In the following, we let $\tb_{\mu_i}^{(0)}$ represent a general zeroth-order multiplier to treat the two series on an equal footing. Equations for the correction amplitudes are determined by requiring $L^{\CCSDT}(\tsbold,\tbzbold,\dtbold,\dtbbold)$ to be stationary with respect to variations in the correction multipliers
\begin{align}
\label{generalampeq}
\frac{\partial L^{\CCSDT}(\tsbold,\tbzbold,\dtbold,\dtbbold)}{\partial \delta \tb_{\mu_i}} = 0 
\qquad
(i=1,2,3)
\end{align}
which reproduces the CCSDT amplitude equations in \ref{ampeq}. It follows that the equations for the correction amplitudes are independent of the choice of zeroth-order multipliers, and the correction amplitudes for the $\ECCn$ and $\CCn$ series are therefore identical to all orders. Equations for the CCSDT multipliers are obtained by requiring $L^{\CCSDT}(\tsbold,\tbzbold,\dtbold,\dtbbold)$ to be stationary with respect to variations in the amplitudes
\begin{align}
\label{generalmulteq}
\frac{\partial L^{\CCSDT}(\tsbold,\tbzbold,\dtbold,\dtbbold)}{\partial \delta t_{\mu_i}} = 0 
\qquad
(i=1,2,3) \ .
\end{align}
Unlike for the correction amplitudes in \ref{ampeq}, the precise form of the multiplier equations will depend upon the choice of zeroth-order multipliers, and the correction multipliers for the $\ECCn$ and $\CCn$ series will therefore be different. To keep this distinction clear, we will refer to the correction multipliers associated with the choices ${\bar{\textbf{t}}}^{(0)} = \zbold$ and ${\bar{\textbf{t}}}^{(0)} = \tbsbold$ 
 as $\dtbbolde$ and $\dtbboldl$, respectively, while the generic notation $\dtbbold$ may refer to either of the series. 
We note that at infinite order, the multipliers---as defined within either the $\ECCn$ or $\CCn$ series---are identical (assuming that both expansions converge), i.e.
\begin{align}
\sum_{n=1}^{\infty} \delta\bar{\textbf{t}}^{E(n)} = 
{^{*}}\bar{\textbf{t}} + \sum_{n=1}^{\infty}\delta\bar{\textbf{t}}^{L(n)} = 
\bar{\textbf{t}}
\end{align}
where $\bar{\textbf{t}}$ is the final set of converged CCSDT multipliers ($\Lambda$-state parameters).\\

To simplify the comparison of the two series, it proves convenient to expand the Lagrangian in \ref{LT1} in a form that emphasizes the dependence on the CCSD multiplier equation in \ref{CCSDmult}
\begin{align}
\nonumber
L^{\CCSDT}(\tsbold,\tbzbold,\dtbold,\dtbbold)  &= 
E^{\CCSD} 
\\
\nonumber
&+ \sum_{i=1}^2 \sum_{\mu_i} \delta t_{\mu_i} 
\bigg( \bra{\hf} [\PhiSD,\hat{\tau}_{\mu_i} ]  \ket{\hf}  + 
\sum_{j=1}^2 \sum_{\nu_j} \tb_{\nu_j}^{(0)}
 \bra{\nu_j}  [\HSD,\hat{\tau}_{\mu_i}]   \ket{\hf} \bigg)
  \\
  \nonumber
  &+ \tfrac{1}{2} \bra{\hf} [ [\PhiSD, \delta\hat{T}_1], \delta\hat{T}_1] \ket{\hf} \\
  \nonumber
&+ \sum_{j=1}^2 \sum_{\nu_j} \tb_{\nu_j}^{(0)}
\bigg( \bra{\nu_j}  [\PhiSD, \delta\hat{T}_3] +  \tfrac{1}{2}[ [\PhiSD, \delta\hat{T}], \delta\hat{T}] + \ldots \ket{\hf} \bigg) \\
\nonumber
&+ \sum_{j=1}^2 \sum_{\nu_j} \delta \tb_{\nu_j}
\bigg( \bra{\nu_j} [\HSD, \delta\hat{T}]  + \tfrac{1}{2}[ [\PhiSD, \delta\hat{T}], \delta\hat{T}] + \ldots \ket{\hf} \bigg)  \\
&+  \sum_{\nu_3} \delta \tb_{\nu_3}
\bigg( \bra{\nu_3} \PhiSD + [\HSD, \delta\hat{T}]  + \tfrac{1}{2}[ [\PhiSD, \delta\hat{T}], \delta\hat{T}] + \ldots \ket{\hf} \bigg)
 \label{LT2}
\end{align}
where we have set $\tzbold=\tsbold$ and used the CCSD amplitude equations in \ref{CCSDamp}.\\

By choosing $\tbzbold = \zbold$ in \ref{LT2}, we arrive at the Lagrangian $L^{\CCSDT}(\tsbold,\zbold,\dtbold,\dtbbolde)$, which reads
\begin{align}
\nonumber
L^{\CCSDT}(\tsbold,\zbold,\dtbold,\dtbbolde)  &= 
E^{\CCSD} 
\\
\nonumber
&+  \bra{\hf} [ \PhiSD, \delta\hat{T}_1 + \delta\hat{T}_2] + \tfrac{1}{2}[[\PhiSD, \delta\hat{T}_1], \delta\hat{T}_1] \ket{\hf}  \\
  \nonumber
&+ \sum_{j=1}^2 \sum_{\nu_j} \tbe_{\nu_j}
\bigg( \bra{\nu_j} [\HSD, \delta\hat{T}]  + \tfrac{1}{2}[ [\PhiSD, \delta\hat{T}], \delta\hat{T}] + \ldots \ket{\hf} \bigg)  \\
&+  \sum_{\nu_3} \tbe_{\nu_3}
\bigg( \bra{\nu_3} \PhiSD + [\HSD, \delta\hat{T}]  + \tfrac{1}{2}[ [\PhiSD, \delta\hat{T}], \delta\hat{T}] + \ldots \ket{\hf} \bigg) \ .
 \label{LT3}
\end{align}
Since the $\dtbbolde$ multipliers multiply the CCSDT amplitude equations in \ref{ampeq}, they may be eliminated from \ref{LT3}, and it follows that $L^{\CCSDT}(\tsbold,\zbold,\dtbold,\dtbbolde)$ equals the CCSDT energy in \ref{ETf3}
\begin{align}
L^{\CCSDT}(\tsbold,\zbold,\dtbold,\dtbbolde)  = E^{\CCSDT} \ .
 \label{LT4}
\end{align}
By evaluating $L^{\CCSDT}(\tsbold,\zbold,\dtbold,\dtbbolde)$ to different orders in the fluctuation potential, we thus arrive at the $\ECCn$ series in \ref{ECCSDTn}, for which the energy corrections may be expressed exclusively in terms of correction amplitudes. The evaluation of the $\ECCn$ energy corrections using \ref{ECCSDTn} corresponds to using the $n+1$ rule, where amplitudes to order $n$ determine the energy to order $n+1$. Alternatively, by exploiting that $L^{\CCSDT}(\tsbold,\zbold,\dtbold,\dtbbolde)$ is variational in the amplitudes as well as the multipliers, it may be evaluated to different orders in the fluctuation potential by using the $2n +1$ and $2n+2$ rules~\cite{helgaker_jorgensen_1988,helgaker_jorgensen_1989,kasper_wigner_rules} for the amplitudes and multipliers (i.e., the amplitudes/multipliers to order $n$ determine the Lagrangian to order $2n+1$/$2n+2$). These two approaches are of course equivalent, but the use of the $2n +1$ and $2n+2$ rules allows for an easy comparison of the $\ECCn$ series to the $\CCn$ series (cf. \ref{appendix_comparison}). \\
% 
%The amplitude equations for $L^{\CCSDT}(\tsbold,\zbold,\dtbold,\dtbbolde)$ are given in Eq.~\eqref{ampeq}, while the multiplier equations are determined by applying the stationary condition in Eq.~\eqref{generalmulteq} to $L^{\CCSDT}(\tsbold,\zbold,\dtbold,\dtbbolde)$. By limiting ourselves to first-order multipliers only, $\tbeone$, these read
%%
%\begin{subequations}
%\label{Emult1}
%\begin{align}
% \tbeone_{\mu_i} &= -
%\epsilon_{\mu_i}^{-1}
% \bra{\hf} [ \PhiSD,\hat{\tau}_{\mu_i}]  \ket{\hf}  
% \qquad (i=1,2) \\
% \tbeone_{\mu_3} &= 0 \ .
%\end{align}
%\end{subequations}
%%
%The lowest-order energy correction of $L^{\CCSDT}(\tsbold,\zbold,\dtbold,\dtbbolde)$ ($E^{(3)}$) may be evaluated in accordance with the $2n+1$ and $2n+2$ rules by expanding $L^{\CCSDT}(\tsbold,\zbold,\dtbold,\dtbbolde)$ in Eq.~\eqref{LT3} in orders of the fluctuation potential and retaining only terms involving first-order amplitudes and multipliers
%%
%\begin{align}
%E^{(3)} = 
%\sum_{i=1}^2 \sum_{\mu_i}  \tbeone_{\mu_i}
% \bra{\mu_i}  [\PhiSD, \delta\hat{T}_3^{(1)}]  \ket{\hf}
% \label{E3alt}
%\end{align}
%%
%where we have used that the first-order singles and doubles correction amplitudes are zero, cf. Eq.~\eqref{amp_pert_expand}. Naturally, the expressions for $E^{(3)}$ obtained from Eq.~\eqref{ECCSDTn2} ($n+1$ rule) and Eq.~\eqref{E3alt} ($2n+1$ and $2n+2$ rules) are identical, as may be verified by an explicit comparison.

By setting $\tbzbold = \tbsbold$ in \ref{LT2}, the resulting Lagrangian $L^{\CCSDT}(\tsbold,\tbsbold,\dtbold,\dtbboldl)$ becomes
\begin{align}
\nonumber
L^{\CCSDT}(\tsbold,\tbsbold,\dtbold,\dtbboldl)  &= 
E^{\CCSD} 
\\
  \nonumber
  &+ \tfrac{1}{2} \bra{\hf} [ [\PhiSD, \delta\hat{T}_1], \delta\hat{T}_1] \ket{\hf} \\
  \nonumber
&+ \sum_{j=1}^2 \sum_{\nu_j} \tbs_{\nu_j}
\bigg( \bra{\nu_j}  [\PhiSD,\delta\hat{T}_3] +  \tfrac{1}{2}  [ [\PhiSD, \delta\hat{T}], \delta\hat{T}] + \ldots \ket{\hf} \bigg) \\
\nonumber
&+ \sum_{j=1}^2 \sum_{\nu_j} \tbl_{\nu_j}
\bigg( \bra{\nu_j} [\HSD, \delta\hat{T}]  + \tfrac{1}{2}  [ [\PhiSD, \delta\hat{T}], \delta\hat{T}] + \ldots \ket{\hf} \bigg)  \\
&+  \sum_{\nu_3} \tbl_{\nu_3}
\bigg( \bra{\nu_3} \PhiSD + [\HSD, \delta\hat{T}]  + \tfrac{1}{2}  [ [\PhiSD, \delta\hat{T}], \delta\hat{T}] + \ldots \ket{\hf} \bigg)
 \label{LT5}
\end{align}
where we have used the CCSD multiplier equations in \ref{CCSDmult}. An expansion of $L^{\CCSDT}(\tsbold,\tbsbold,\dtbold,\dtbboldl)$ in orders of the fluctuation potential defines the recently proposed $\CCn$ series~\cite{eriksen2014lagrangian}. The $\ECCn$ series begins at third order (see \ref{ECCSDTn}), while the $\CCn$ series starts already at second order. The series are evidently different, and in \ref{sec:comparison} we perform an explicit comparison of them.

%
%As for the $\ECCn$ series, the equations for the correction amplitudes are given in Eq.~\eqref{amp_pert_expand}, while the multiplier equations are obtained, as in Section \ref{subsub:eccsdt}, by applying Eq.~\eqref{generalmulteq} to the $L^{\CCSDT}(\tsbold,\tbsbold,\dtbold,\dtbboldl)$ Lagrangian. Again, limiting ourselves to first-order multipliers only, $\tblone$, these read
%%
%\begin{subequations}
%\label{Lmult1}
%\begin{align}
%\tblone_{\mu_i} &= 0 \qquad (i=1,2) \\
%\tblone_{\mu_3} &= - \epsilon_{\mu_3}^{-1} \sum_{i=1}^2 \tbs_{\mu_i} \bra{\mu_i} [ \PhiSD,\hat{\tau}_{\mu_3} ] \ket{\hf} \ . 
%\end{align}
%\end{subequations}
%%
%From Eq.~\eqref{LT5}, it follows that the $\CCn$ series starts at second order, and the lowest-order energy, $L^{(2)}$ (evaluated using the $2n +1$ and $2n+2$ rules), is given by
%%
%\begin{align}
%\label{L2}
%L^{(2)} &= \sum_{i=1}^2 \sum_{\mu_i} \tbs_{\mu_i} \bra{\mu_i} [ \PhiSD, \delta\hat{T}_3^{(1)} ] \ket{\hf} \ . 
%\end{align}
%%
%

%
%%%%%%%%%%%%%%%%%%%%%%%%%%%%%%%%%%%%%%%%%%%%%%%%%%%%%%%%%%%%%%%%%%%%
%                                                                    	Comparison of the $\ECCn$ and $\CCn$ series
%%%%%%%%%%%%%%%%%%%%%%%%%%%%%%%%%%%%%%%%%%%%%%%%%%%%%%%%%%%%%%%%%%%%
%
\subsection{Comparison of the $\ECCn$ and $\CCn$ series}\label{sec:comparison}

In this section, we compare the $\ECCn$ and $\CCn$ series---first, we discuss the origin of the difference between the series from a formal point of view, and next, we compare the two lowest-order multipliers and energy corrections for the two series.

As shown in \ref{sec:general}, both series can be derived from \ref{LT2} with different choices of parent multipliers, resulting in the Lagrangians, $L^{\CCSDT}(\tsbold,\zbold,\dtbold,\dtbbolde)$ and $L^{\CCSDT}(\tsbold,\tbsbold,\dtbold,\dtbboldl)$, for the $\ECCn$ and $\CCn$ series, respectively. The $L^{\CCSDT}(\tsbold,\zbold,\dtbold,\dtbbolde)$ Lagrangian reduces to the standard CCSDT energy expression, because no zeroth-order multipliers enter the Lagrangian and because the correction multipliers, $\delta \tb$, are multiplied by the CCSDT amplitude equations (cf. \ref{LT3} and \ref{LT4}). However, the $L^{\CCSDT}(\tsbold,\tbsbold,\dtbold,\dtbboldl)$ Lagrangian cannot be subject to a similar reduction, since certain terms in the CCSDT amplitude equations were cancelled when the CCSD multiplier equations were used to manipulate \ref{LT2} to arrive at \ref{LT5}. Consequently, the CCSD multipliers cannot be removed from \ref{LT5}, and $L^{\CCSDT}(\tsbold,\tbsbold,\dtbold,\dtbboldl)$ can therefore not be reduced to the standard CCSDT energy expression.

In the same way as the CCSD amplitudes formally depend on the fluctuation potential to infinite order in the space of all single and double excitations, so do the CCSD multipliers (or CCSD $\Lambda$-state parameters). Thus, since the CCSD multipliers are not counted as zeroth-order parameters in the $\ECCn$ series (unlike in the $\CCn$ series), orders are necessarily counted differently in the perturbative expansions of $L^{\CCSDT}(\tsbold,\tbsbold,\dtbold,\dtbboldl)$ and $L^{\CCSDT}(\tsbold,\zbold,\dtbold,\dtbbolde)$; however, we may compare the two series by comparing their leading-order, next-to-leading-order, etc., corrections to one another, which we will do numerically in \ref{sec:results}. 

On that note, we have theoretically compared the lowest- and next-to-lowest-order corrections of the two series in \ref{appendix_comparison}. In summary, we find that for the $L^{\CCSDT}(\tsbold,\tbsbold,\dtbold,\dtbboldl)$ Lagrangian, a good zeroth-order description of the CCSDT $\Lambda$-state in the space of all single and double excitations is used (the CCSD $\Lambda$-state), and the lowest-order multiplier correction therefore occurs in the triples space (cf. \ref{CCSDmult_order_exp_3_1}). For the $L^{\CCSDT}(\tsbold,\zbold,\dtbold,\dtbbolde)$ Lagrangian, on the other hand, there is no representation of the CCSDT $\Lambda$-state at zeroth order, and the leading-order contributions to this state (in the form of the first-order multipliers, $\tbeone_{\mu_i}$) hence occur within the singles and doubles space, while triples multipliers first enter the $\ECCn$ series at the following (second) order (cf. \ref{CCSDmult_order_exp_2}). In fact, we find that the first- and second-order singles and doubles multipliers, $\{ \tbeone_{\mu_i}, \tbetwo_{\mu_i} \}$ for $i=1,2$, are nothing but the two lowest-order contributions to the CCSD $\Lambda$-state parameters, if the CCSD multiplier equations in \ref{CCSDmult} are solved perturbatively (cf. \ref{tbs_pert_expansion}). Furthermore, the lowest-order $\ECCn$ and $\CCn$ energy corrections are found to have the same structural form, however, they are expressed in terms of different sets of multipliers (cf. \ref{E3alt_plus_E4alt_app} and \ref{L2alt_plus_L3alt_app}). Thus, the $\ECCn$ series may be viewed as attempting to compensate for the poor (non-existing) guess at the CCSDT $\Lambda$-state by mimicking the $\CCn$ series as closely as possible within a perturbational framework. For both series, a perturbative solution of the CCSDT $\Lambda$-state is embedded into the energy corrections, and the $\ECCn$ series is thus trailing behind the $\CCn$ series from the onset of the perturbation expansion. Again, this motivates the claim that it is advantageous to consider the CC energy as the stationary point of an energy functional in both the CC and $\Lambda$-state parameters, and hence that perturbative expansions are optimally carried out whenever the two states are treated on an equal footing~\cite{arponen_ann_phys_1983,arponen_pra_1987,kvaal_jcp_2012,kvaal_mol_phys_2013}. For higher orders in the perturbation, a direct comparison of the $\ECCn$ and $\CCn$ series is more intricate, but we note that they will differ to all orders. Only in the infinite-order limit are the two bound to agree ($L^{\CCSDT}(\tsbold,\zbold,\dtbold,\dtbbolde)$ and $L^{\CCSDT}(\tsbold,\tbsbold,\dtbold,\dtbboldl)$ both equal the CCSDT energy), but this is a natural consequence of the fact that the CC energy may be described in terms of fully converged CC amplitudes alone.\\

In conclusion, the $L^{\CCSDT}(\tsbold,\zbold,\dtbold,\dtbbolde)$ and $L^{\CCSDT}(\tsbold,\tbsbold,\dtbold,\dtbboldl)$ Lagrangians have the same parent energy (CCSD) and the same target energy (CCSDT), and all contributions of both series will be trivially term-wise size extensive to all orders, as they are all expressed in terms of (linked) commutator expressions. However, the paths between these two CC energies, as defined by a perturbation expansion, are obviously very different for the two series. Conceptually, the expansion point for the $\ECCn$ series is formally the {\it CCSD energy} (only the CCSD amplitude equations are satisfied), while the expansion point for the $\CCn$ series is the {\it CCSD Lagrangian} (the CCSD amplitude {\it and} CCSD multiplier equations are satisfied). For the energy-based $\ECCn$ series, the Lagrangian is thus merely a mathematical tool that allows for correction energies to be obtained using amplitude and multiplier corrections that satisfy the $2n+1$ and $2n+2$ rules. On the contrary, the $\CCn$ series is deeply rooted within a bivariational Lagrangian formulation and has no energy-based analogue.

%
%%%%%%%%%%%%%%%%%%%%%%%%%%%%%%%%%%%%%%%%%%%%%%%%%%%%%%%%%%%%%%%%%%%
%                                                                    				Numerical results
%%%%%%%%%%%%%%%%%%%%%%%%%%%%%%%%%%%%%%%%%%%%%%%%%%%%%%%%%%%%%%%%%%%
%
\section{Numerical results}\label{sec:results}

The performance of the models of the $\CCn$ series has recently been theoretically as well as numerically compared to a variety of alternative triples models for two sets of closed-shell~\cite{eriksen2015convergence} and open-shell species~\cite{open_shell_triples_eriksen_2015,open_shell_quadruples_eriksen_2016}, and the formal convergence of the series (through sixth order in the perturbation) has been confirmed. Furthermore, the performance of the higher-order $\CCn$ models with respect to the target CCSDT model was found to be essentially independent of the HF reference used, and thus, independent of the spin of the ground state. In this section, we assess the numerical performance of the $\ECCn$ models (once again measured against results obtained with the target CCSDT model) in order to compare the rate of convergence throughout the series to that of the $\CCn$ series.

We here use the two test sets previously used in Refs. \citenum{eriksen2015convergence}, \citenum{quadruples_pert_theory_jcp_2015}, \citenum{open_shell_triples_eriksen_2015}, and \citenum{open_shell_quadruples_eriksen_2016}: $\textbf{(i)}$ 17 closed-shell molecules, all optimized at the all-electron CCSD(T)/cc-pCVQZ level of theory, and $\textbf{(ii)}$ 18 open-shell atoms and radicals, all optimized at the all-electron CCSD(T)/cc-pVQZ level of theory. For a specification of the members of the closed- and open-shell test sets as well as tabularized molecular geometries, cf. Refs. \citenum{mest} and the papers describing the HEAT thermochemical model~\cite{tajti2004heat,bomble2006high,harding2008high}, respectively. All of the closed-shell calculations are based on a restricted HF (RHF) reference, while unrestricted HF (UHF) as well as restricted open-shell HF (ROHF) trial functions have been used for the open-shell calculations. The correlation-consistent cc-pVTZ basis set~\cite{dunning_1} is used throughout for all of the reported valence-electron (frozen-core) results, and the Aquarius program~\cite{aquarius} has been used for all of the calculations.

\begin{figure}
        \centering
        \begin{subfigure}[b]{0.47\textwidth}
                \includegraphics[width=\textwidth,bb=0 0 488 378]{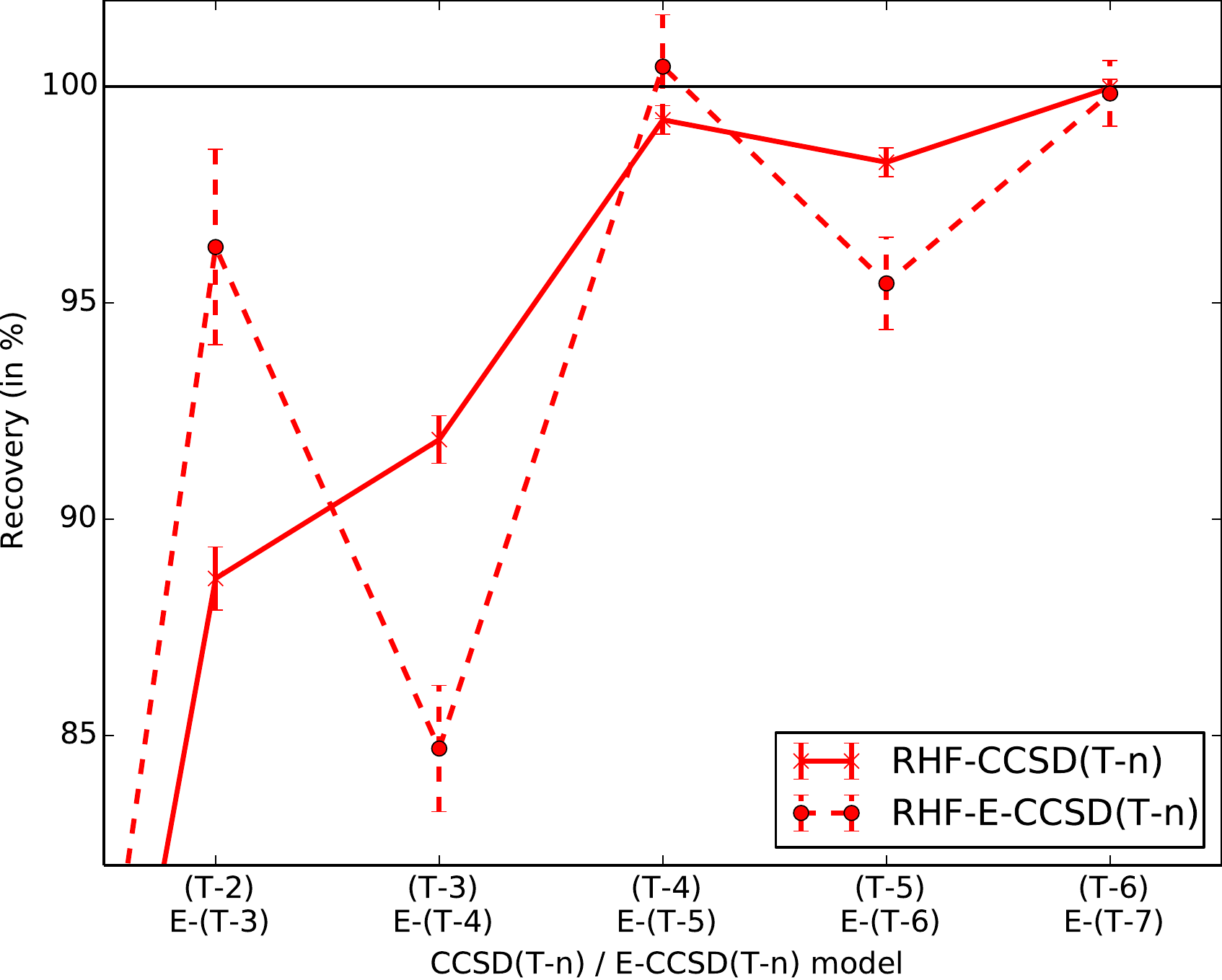}
                \caption{RHF recoveries}
                \label{e_t_n_recoveries_rhf_figure}
        \end{subfigure}
        \vspace{0.4cm}
        \hspace{0.4cm}
        \begin{subfigure}[b]{0.47\textwidth}
                \includegraphics[width=\textwidth,bb=0 0 488 378]{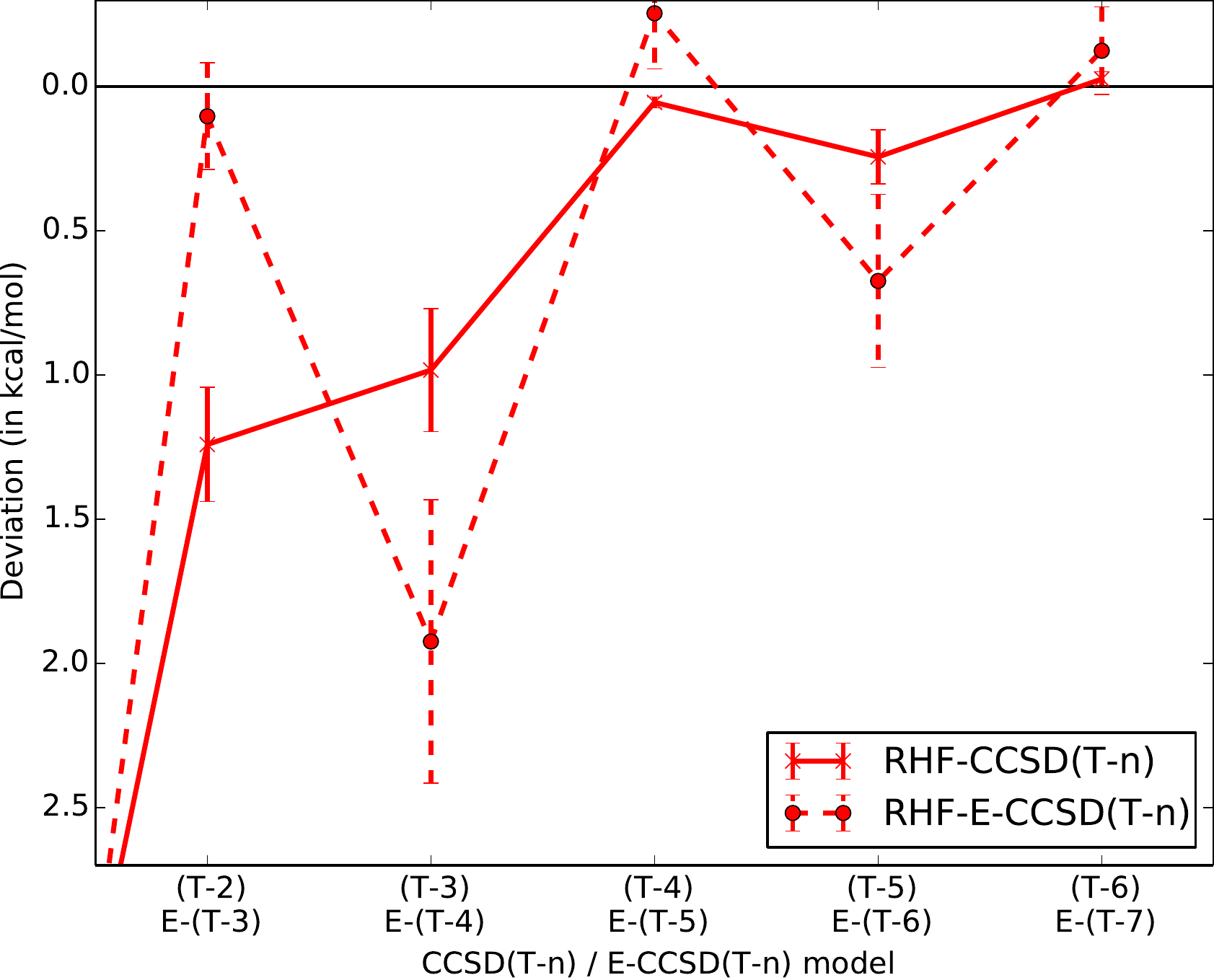}
                \caption{RHF deviations}
                \label{e_t_n_abs_diff_rhf_figure}
        \end{subfigure}
        \begin{subfigure}[b]{0.47\textwidth}
                \includegraphics[width=\textwidth,bb=0 0 488 378]{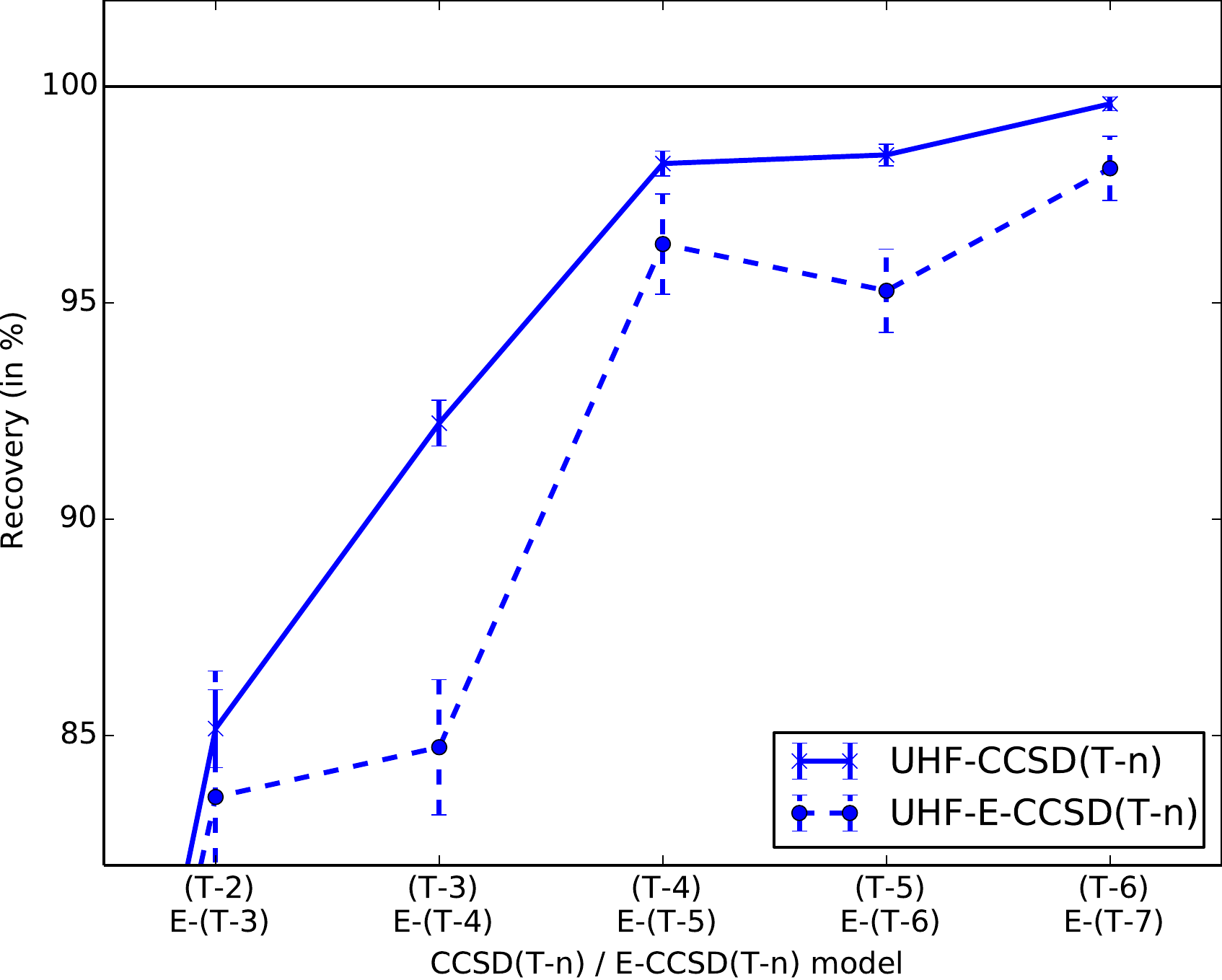}
                \caption{UHF recoveries}
                \label{e_t_n_recoveries_uhf_figure}
        \end{subfigure}
        \vspace{0.4cm}
        \hspace{0.4cm}
        \begin{subfigure}[b]{0.47\textwidth}
                \includegraphics[width=\textwidth,bb=0 0 488 378]{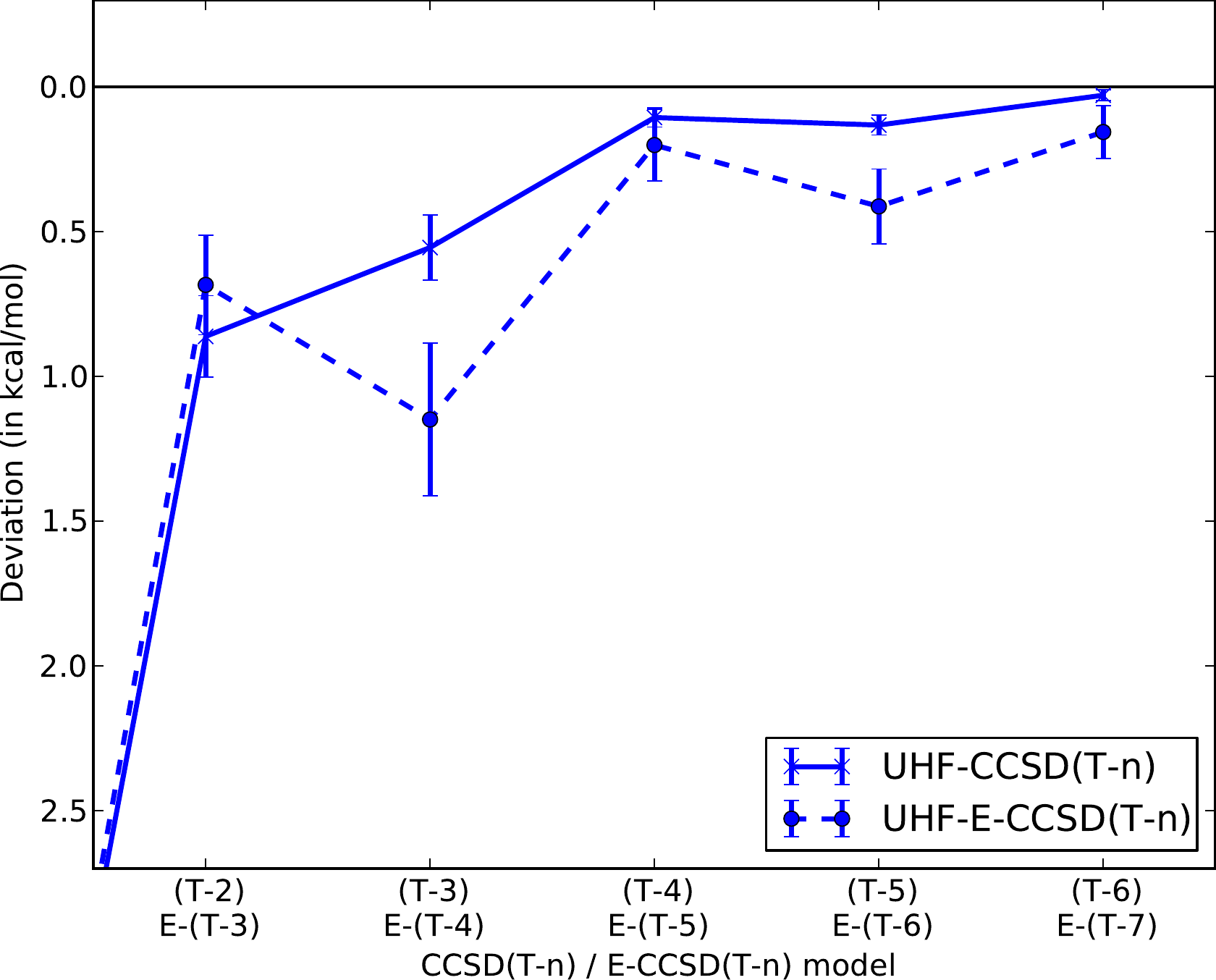}
                \caption{UHF deviations}
                \label{e_t_n_abs_diff_uhf_figure}
        \end{subfigure}
        \begin{subfigure}[b]{0.47\textwidth}
                \includegraphics[width=\textwidth,bb=0 0 488 378]{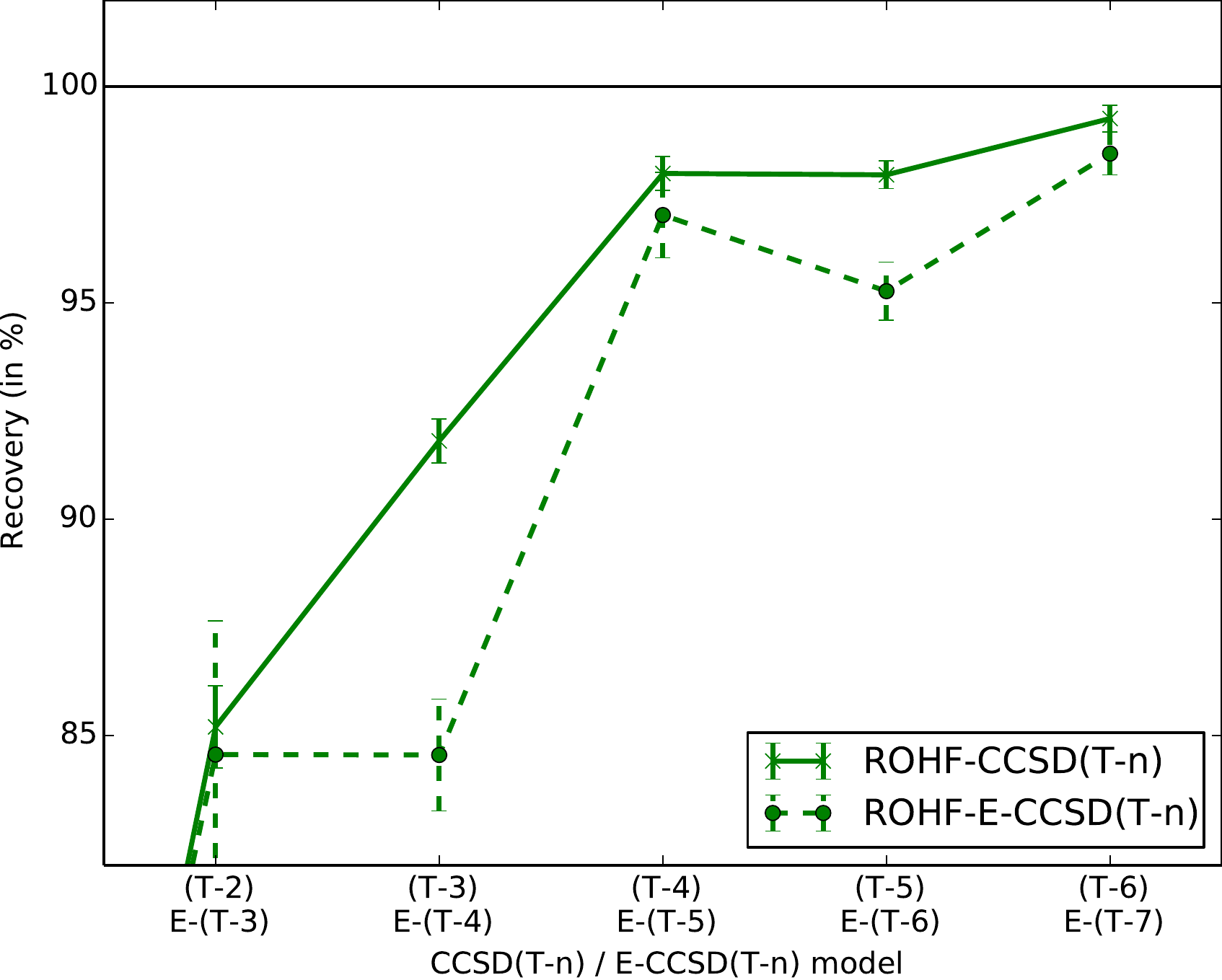}
                \caption{ROHF recoveries}
                \label{e_t_n_recoveries_rohf_figure}
        \end{subfigure}
        \hspace{0.4cm} 
        \begin{subfigure}[b]{0.47\textwidth}
                \includegraphics[width=\textwidth,bb=0 0 488 378]{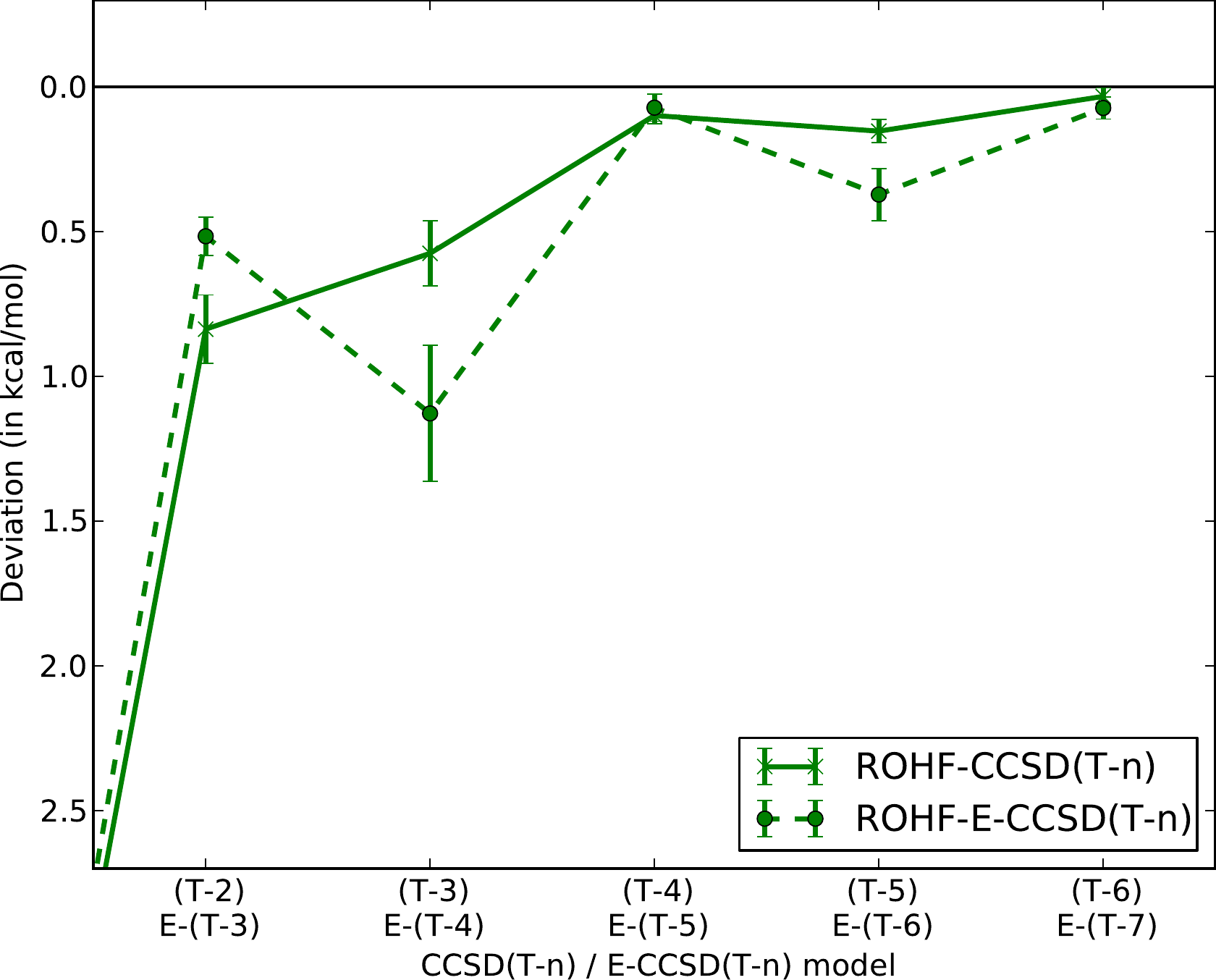}
                \caption{ROHF deviations}
                \label{e_t_n_abs_diff_rohf_figure}
        \end{subfigure}
        \caption{Mean recoveries of (in $\%$, \ref{e_t_n_recoveries_rhf_figure}, \ref{e_t_n_recoveries_uhf_figure}, and \ref{e_t_n_recoveries_rohf_figure}) and mean deviations from (in kcal/mol, \ref{e_t_n_abs_diff_rhf_figure}, \ref{e_t_n_abs_diff_uhf_figure}, and \ref{e_t_n_abs_diff_rohf_figure}) the triples energy $E^{\text{T}}$ for the $\CCn$ and $\ECCn$ series using 
         RHF, UHF, and ROHF references.
       The error bars show the standard error of the mean.
       }
        \label{e_t_n_figure}
\end{figure}

In \ref{e_t_n_figure}, we consider the performance of the five lowest-order models of the $\ECCn$ and $\CCn$ hierarchies. Mean recoveries (in $\%$) of the triples correlation energy, $E^{\text{T}} = E^{\CCSDT} - E^{\CCSD}$, are presented in \ref{e_t_n_recoveries_rhf_figure}, \ref{e_t_n_recoveries_uhf_figure}, and \ref{e_t_n_recoveries_rohf_figure}, while mean deviations from $E^{\text{T}}$ (in kcal/mol) are presented in \ref{e_t_n_abs_diff_rhf_figure}, \ref{e_t_n_abs_diff_uhf_figure}, and \ref{e_t_n_abs_diff_rohf_figure}. In all cases, we report statistical error measures generated from the individual results, cf. the supplementary material~\bibnote{See supplementary material at [AIP URL] for individual recoveries and deviations. Individual $\CCn$ results are given in Ref. \citenum{open_shell_triples_eriksen_2015}.}. As noted in \ref{sec:comparison}, the $\ECCn$ and $\CCn$ series start at third and second order, respectively, but we may group these together like $\ECCthree$/$\CCtwo$, $\ECCfour$/$\CCthree$, etc.

The results in \ref{e_t_n_figure} show that the $\CCn$ models in general yield smaller mean and standard errors than their $\ECCn$ counterparts, and the $\CCn$ series furthermore exhibits a more stable convergence than the $\ECCn$ series. In other words, the rate of convergence is improved in the $\CCn$ series over the $\ECCn$ series. For most of the considered molecules, the $\ECCn$ corrections are negative/positive for uneven/even orders, which leads to the oscillatory convergence behavior observed for the $\ECCn$ series in \ref{e_t_n_figure}. Some oscillatory behavior is also observed for the $\CCn$ series, but this is much less prominent, and primarily observed beyond fourth order. 
%Since the lowest-order multipliers of the $\ECCn$ series represent the lowest-order contributions to a perturbative solution of the CCSD multiplier equations
%(see Section~\ref{sec:comparison} and Appendix~\ref{appendix_comparison}), we argue that the oscillatory nature of the $\ECCn$ results may be connected to the occasional oscillatory convergence through the first few iterations of the $\Lambda$-equations (using the standard Jacobi procedure without convergence acceleration).
 The superior stability of the $\CCn$ series compared to the $\ECCn$ series is also manifested in the smaller standard deviations for the former (in \ref{e_t_n_figure} represented in terms of standard errors of the mean). Some of the molecules, however, differ considerably from the mean trends in \ref{e_t_n_figure}. For example, methylene (CH$_2$) and ozone (O$_3$) are notoriously difficult cases due to significant multireference character\cite{mest,eriksen2015convergence}, and, for both, we observe a rather slow convergence throughout either of the series. While for O$_3$, the convergence towards the CCSDT limit is oscillatory, for CH$_2$, the convergence is stabile, yet slow, cf. Table S4 of the supplementary material for the $\ECCn$ results. Similar, but significantly less pronounced problems, are observed for the $\CCn$ series~\cite{open_shell_triples_eriksen_2015}. Finally, we note how the results obtained using the two open-shell references (UHF and ROHF) are similar, and also that the general behavior for the mean deviations are similar to the RHF results.

From an application point of view, the $\CCn$ series is practically converged onto the CCSDT limit at the $\CCfour$ model (robust for closed- and open-shell systems), while two additional corrections (two additional orders in the perturbation) are needed in the $\ECCn$ series in order to match these results (the $\ECCseven$ model). However, even for the $\ECCseven$ model, the standard deviations (for both recoveries and deviations) are larger than for the $\CCfour$ model. If results more accurate than those provided by the $\CCfour$/$\ECCseven$ models are desired, it is in general necessary to also account for the effects of quadruple excitations, as the quadruples energy contribution may easily exceed the difference between the CCSDT and $\CCfour$ energies. In such cases, the recently proposed CCSDT(Q--$n$) models~\cite{eriksen2014lagrangian,quadruples_pert_theory_jcp_2015} may offer attractive alternatives to the iterative CCSDTQ model. These models are theoretically on par with the $\CCn$ models, but expand the CCSDTQ--CCSDT energy difference, rather than the CCSDT--CCSD difference, in orders of the MP fluctuation potential.

In conclusion, a number of similarities exist between the $\ECCn$ and $\CCn$ series, but both the magnitude of the (individual) errors as well as the oscillatory convergence pattern are significantly reduced in the $\CCn$ series, as compared to the $\ECCn$ series. This improvement is in line with the theoretical analysis in \ref{sec:comparison}. More information about the expansion point is used in the $\CCn$ series, where the CCSD amplitudes and CCSD multipliers are both built into the perturbative corrections, to yield a faster and more balanced rate of convergence than that observed for the $\ECCn$ series, in which only the CCSD amplitudes are used to construct the energy corrections. Based on the results for the five lowest-order models in \ref{e_t_n_figure}, the $\ECCn$ and $\CCn$ series both appear to converge for all of the considered molecules, although slowly for some of the notoriously difficult cases. However, a formal convergence analysis is required to firmly establish whether the series indeed converge (i.e., establish the radius of convergence for the series). This will be the subject of a forthcoming paper.

%
%%%%%%%%%%%%%%%%%%%%%%%%%%%%%%%%%%%%%%%%%%%%%%%%%%%%%%%%%%%%%%%%%%%
%                                                                    				Conclusion and perspectives
%%%%%%%%%%%%%%%%%%%%%%%%%%%%%%%%%%%%%%%%%%%%%%%%%%%%%%%%%%%%%%%%%%%
%
\section{Conclusion and perspectives}\label{sec:conclusion}

We have developed the $\ECCn$ perturbation series and compared it to the recently proposed $\CCn$ series in order to gain new insights into the importance of treating amplitudes and multipliers (parameters of the $\Lambda$-state) on an equal footing whenever perturbation expansions are developed within CC theory. Both series represent a perturbation expansion of the difference between the CCSD and CCSDT energies, and they share the same common set of correction amplitudes. The $\ECCn$ series formally describes an expansion around the CCSD energy point (CCSD amplitude equations are satisfied), while the $\CCn$ series may be viewed as an expansion around the CCSD Lagrangian point (CCSD amplitude equations {\it and} CCSD multiplier equations are satisfied). The two series are therefore different, and the $\CCn$ series is found to converge more rapidly towards the CCSDT target energy, since all available information at the CCSD expansion point is utilized.

The presented analysis may be generalized to any perturbation expansion representing the difference between a parent CC model and a higher-level target CC model. 
For developments of CC perturbation expansions, we thus generally advocate the use a bivariational Lagrangian CC formulation to ensure an optimal rate of convergence in terms of term-wise size extensive corrections towards the target energy. For example, two perturbation series formulated around the CCSDT energy (E-CCSDT(Q--$n$)) and CCSDT Lagrangian (CCSDT(Q--$n$)) expansion points, respectively, to describe an expansion towards the CCSDTQ target energy in order the M{\o}ller-Plesset fluctuation potential, are also bound to exhibit different rates of convergence, following a similar line of arguments.\\

In quantum chemistry, a Lagrangian energy functional has traditionally been viewed merely as a convenient mathematical tool for deriving perturbative expansions, however, one that would give rise to expansions that are identical to those based on the standard energy. The present work highlights how this equivalence between energy- and Lagrangian-based perturbation theory only holds whenever the zeroth-order parameters do not depend on the perturbing operator, as is, for example, the case for standard MP perturbation theory where the zeroth-order parameters vanish. Thus, when the zeroth-order parameters are independent of the perturbing operator, a Lagrangian formulation is merely of mathematical convenience, but, for perturbation-dependent zeroth-order parameters (e.g., like those of the right- (CC) and left-hand ($\Lambda$) eigenstates of a non-Hermitian CC similarity-transformed Hamiltonian), a bivariational Lagrangian formulation is in general expected to lead to a faster and more stable convergence than a corresponding energy formulation. This is an important point to keep in mind for future developments and applications involving perturbation expansions.

%
%%%%%%%%%%%%%%%%%%%%%%%%%%%%%%%%%%%%%%%%%%%%%%%%%%%%%%%%%%%%%%%%%%%
%                                                                    			     ACKNOWLEDGMENT
%%%%%%%%%%%%%%%%%%%%%%%%%%%%%%%%%%%%%%%%%%%%%%%%%%%%%%%%%%%%%%%%%%%
%
\section*{Acknowledgments}

K. K., J. J. E., and P. J. acknowledge support from the European Research Council under the European Union's Seventh Framework Programme (FP/2007-2013)/ERC Grant Agreement No. 291371. J. O. acknowledges support from the Danish Council for Independent Research, DFF-4181-00537, and D. A. M. acknowledges support from the US National Science Foundation (NSF) under grant number ACI-1148125/1340293.

\newpage

%
%%%%%%%%%%%%%%%%%%%%%%%%%%%%%%%%%%%%%%%%%%%%%%%%%%%%%%%%%%%%%%%%%%%
%                                                                    			     	APPENDIX
%%%%%%%%%%%%%%%%%%%%%%%%%%%%%%%%%%%%%%%%%%%%%%%%%%%%%%%%%%%%%%%%%%%
%
\appendix
\section{Explicit lowest-order $\ECCn$ and $\CCn$ correction energies}\label{appendix_comparison}

In the present appendix, we compare the lowest- and next-to-lowest-order corrections of the $\ECCn$ and $\CCn$ series. For this, we need a closed-form expression for the CCSDT multipliers (i.e., the CCSDT $\Lambda$-state parameters), which, from \ref{generalmulteq}, reads
\begin{align}
\label{CCSDT_mult_explicit}
\bar{t}_{\mu_i} = -\epsilon^{-1}_{\mu_i}\big(\bra{\hf} [ \Phi^{\hat{T}}, \hat{\tau}_{\mu_i} ] \ket{\hf} + \sum_{j=1}^3 \sum_{\nu_j} \bar{t}_{\nu_j}\bra{\nu_j} [ \Phi^{\hat{T}}, \hat{\tau}_{\mu_i} ] \ket{\hf}\big)
\end{align}
where we will again partition the CCSDT cluster operator, $\hat{T}$, as $\hat{T} = {^{\ast}}\hat{T} + \delta\hat{T}$. \ref{CCSDT_mult_explicit} may now be expanded in orders of the fluctuation potential (cf. \ref{multpara})
\begin{align}
\bar{\textbf{t}} = \bar{\textbf{t}}^{(0)} + \delta\bar{\textbf{t}}^{(1)} + \delta\bar{\textbf{t}}^{(2)} + \ldots \label{CCSDT_mult_order_exp_1}
\end{align}
If $\bar{\textbf{t}}^{(0)} = \bm{0}$, the multiplier corrections in \ref{CCSDT_mult_order_exp_1} will be those belonging to the $\ECCn$ series, and the two lowest-order corrections are given by
\begin{subequations}
\label{CCSDmult_order_exp_2}
\begin{align}
\tbeone_{\mu_i} &= -\epsilon^{-1}_{\mu_i}\bra{\hf} [ \PhiSD, \hat{\tau}_{\mu_i} ] \ket{\hf} \label{CCSDmult_order_exp_2_1} \\
\tbetwo_{\mu_i} &= -\epsilon^{-1}_{\mu_i}\sum_{j=1}^2 \sum_{\nu_j} \tbeone_{\nu_j}\bra{\nu_j} [ \PhiSD, \hat{\tau}_{\mu_i} ] \ket{\hf} \ . \label{CCSDmult_order_exp_2_2}
\end{align}
\end{subequations}
It follows that the $\ECCn$ series has non-vanishing first-order multipliers only in the singles and doubles space ($\tbeone_{\mu_3}=0$), and second-order multipliers for all excitation levels ($i=1,2,3$). 

Using \ref{ECCSDTn}, the two leading-order corrections of the $\ECCn$ series may be evaluated using the $n+1$ rule for the amplitudes
\begin{subequations}
\label{E3alt_E4alt_app_n_1}
\begin{align}
E^{(3)} &= \sum^2_{j=1}  \bra{\hf} [\PhiSD, \delta\hat{T}_j^{(2)}]  \ket{\hf} \label{E3alt_app_n_1} \\
E^{(4)} &= \sum^2_{j=1} \bra{\hf} [\PhiSD, \delta\hat{T}_j^{(3)}]  \ket{\hf}  \ .
\label{E4alt_app_n_1}
%&= \sum_{i=1}^2 \sum_{\mu_i}  \tbeone_{\mu_i}\bra{\mu_i}  [\PhiSD, \delta\hat{T}_3^{(1)}]  \ket{\hf} \label{E3alt_app}
\end{align}
\end{subequations}
Alternatively, using the Lagrangian in \ref{LT3} and the $2n+1$/$2n+2$ rules\cite{helgaker_jorgensen_1988,helgaker_jorgensen_1989,kasper_wigner_rules} for the amplitudes/multipliers,
$E^{(3)}$ and $E^{(4)}$ may be written as
\begin{subequations}
\label{E3alt_E4alt_app_2n_1}
\begin{align}
E^{(3)} &= \sum_{i=1}^2 \sum_{\mu_i}  \tbeone_{\mu_i}\bra{\mu_i}  [\PhiSD, \delta\hat{T}_3^{(1)}]  \ket{\hf} \label{E3alt_app_2n_1} \\
E^{(4)} &= \sum^3_{k=1} \big\{\sum_{i=1}^2 \sum_{\mu_i} \tbeone_{\mu_i}\bra{\mu_i} [\PhiSD, \delta\hat{T}_k^{(2)}]  \ket{\hf} \big\} \ . \label{E4alt_app_2n_1}
%&= \sum_{l=1}^3 \sum_{\mu_l}  \tbetwo_{\mu_l}\bra{\mu_l}  [\PhiSD, \delta\hat{T}_3^{(1)}]  \ket{\hf} \ . \label{E4alt_app}
\end{align}
\end{subequations}
The expressions in \ref{E3alt_E4alt_app_n_1} and \ref{E3alt_E4alt_app_2n_1} are of course  equivalent as may be verified by explicit comparison. Finally, the $E^{(4)}$ energy in \ref{E4alt_app_2n_1} may be further recast by expanding the second-order correction amplitudes, $\delta\textbf{t}^{(2)}$, given in \ref{amp_pert_expand}
\begin{align}
E^{(4)} &= -\sum^3_{k=1}\sum_{\nu_k}\big\{\epsilon^{-1}_{\nu_k}\sum_{i=1}^2 \sum_{\mu_i} \tbeone_{\mu_i}\bra{\mu_i} [\PhiSD, \hat{\tau}_{\nu_k}] \ket{\hf} \bra{\nu_k} [\PhiSD,\delta\hat{T}^{(1)}_{3}] \ket{\hf} \big\} \nonumber \\
&= \sum_{k=1}^3 \sum_{\nu_k}  \tbetwo_{\nu_k}\bra{\nu_k}  [\PhiSD, \delta\hat{T}_3^{(1)}]  \ket{\hf} \ . \label{E4alt_app_final}
\end{align}
By taking the sum of the third- and fourth-order $\ECCn$ energies in \ref{E3alt_app_n_1} and \ref{E4alt_app_final}, respectively, we can write
\begin{align}
E^{(3)} + E^{(4)} &= \sum_{i=1}^2 \sum_{\mu_i} \big(\tbeone_{\mu_i} + \tbetwo_{\mu_i} \big)\bra{\mu_i}  [\PhiSD, \delta\hat{T}_3^{(1)}]  \ket{\hf} \nonumber \\
&\phantom{=} \ + \sum_{\mu_3}\tbetwo_{\mu_3} \bra{\mu_3}  [\PhiSD, \delta\hat{T}_3^{(1)}]  \ket{\hf} \ .\label{E3alt_plus_E4alt_app}
\end{align}
To evaluate the two lowest-order energy corrections of the $\CCn$ series ($\bar{\textbf{t}}^{(0)} = \tbsbold$), we only need to consider the first-order multipliers, which read\cite{eriksen2014lagrangian}
\begin{subequations}
\label{CCSDmult_order_exp_3_1}
\begin{align}
\tblone_{\mu_1} &= \tblone_{\mu_2} = 0 \\
\tblone_{\mu_3} &= -\epsilon^{-1}_{\mu_3}\sum^2_{j=1}{^{\ast}}\bar{t}_{\nu_j}\bra{\nu_j} [ \PhiSD, \hat{\tau}_{\mu_3} ] \ket{\hf} \ . \label{CCSDmult_order_exp_3_1trip}
\end{align}
\end{subequations}
By applying the $2n+1$/$2n+2$ rules to \ref{LT5}, the two leading corrections of the $\CCn$ series are given as
\begin{align}
L^{(2)} = \sum_{i=1}^2 \sum_{\mu_i} \tbs_{\mu_i} \bra{\mu_i} [ \PhiSD, \delta\hat{T}_3^{(1)} ] \ket{\hf} \label{L2alt_app}
\end{align}
and
\begin{align}
L^{(3)} = \sum_{\mu_3} \tblone_{\mu_3}\bra{\mu_3} [\PhiSD, \delta\hat{T}_3^{(1)}]  \ket{\hf} \label{L3alt_app}
\end{align}
the sum of which becomes
\begin{align}
L^{(2)} + L^{(3)} &= \sum_{i=1}^2 \sum_{\mu_i} \tbs_{\mu_i} \bra{\mu_i} [ \PhiSD, \delta\hat{T}_3^{(1)} ] \ket{\hf} \nonumber \\
&\phantom{=} \ + \sum_{\mu_3} \tblone_{\mu_3}\bra{\mu_3} [\PhiSD, \delta\hat{T}_3^{(1)}]  \ket{\hf} \ . \label{L2alt_plus_L3alt_app}
\end{align}

We may now compare the energy sums in \ref{E3alt_plus_E4alt_app} and \ref{L2alt_plus_L3alt_app}. Since the two leading-order multiplier corrections of the $\ECCn$ series are independent of the triple excitations in the CCSDT ansatz, these will equal the two lowest-order contributions to the CCSD $\Lambda$-state parameters, i.e.
\begin{equation}
\label{tbs_pert_expansion}
\tbs_{\mu_i} = \tbeone_{\mu_i} + \tbetwo_{\mu_i} + \mathcal{O}(3)
\qquad (i=1,2)
\end{equation}
where $\mathcal{O}(3)$ denotes terms of third and higher orders in the fluctuation potential.
 For this reason, the first term on the right-hand side of \ref{E3alt_plus_E4alt_app} may be viewed as mimicking the first term on the right-hand side of \ref{L2alt_plus_L3alt_app}, and the same applies for the second term on the right-hand side of the two equations, by noticing that the $\bar{\textbf{t}}^{E(2)}_3$ multipliers of \ref{CCSDmult_order_exp_2_2} are similar to the $\bar{\textbf{t}}^{L(1)}_3$ multipliers of \ref{CCSDmult_order_exp_3_1trip}, with the notable exception that $\bar{\textbf{t}}^{E(1)} \rightarrow {^{\ast}}\bar{\textbf{t}}$ in moving from the $\ECCn$ to the $\CCn$ series.

\newpage

%
%\bibliography{biblio}
%
\providecommand*\mcitethebibliography{\thebibliography}
\csname @ifundefined\endcsname{endmcitethebibliography}
  {\let\endmcitethebibliography\endthebibliography}{}

\end{document}